\documentclass[10pt, oneside]{article}   	
\usepackage{geometry}                		
\pdfoutput=1
\usepackage{graphicx}				
\title{MOSCAB: A geyser-concept bubble chamber to be used in a dark matter search}
\begin{document}
\maketitle
The MOSCAB Collaboration:
A.~Antonicci$\mathrm{^1}$,
M.~Ardid$\mathrm{^2}$,
R.~Bertoni$\mathrm{^1}$,
G.~Bruno$\mathrm{^3}$,
N.~Burgio$\mathrm{^4}$,
G.~Caruso$\mathrm{^5}$,
D.~Cattaneo$\mathrm{^6}$,
F.~Chignoli$\mathrm{^{1,7}}$,
M.~Clemenza$\mathrm{^{1,7}}$,
M.~Corcione$\mathrm{^{5,8,b}}$,
L.~Cretara$\mathrm{^{5}}$,
D.~Cundy$\mathrm{^{7,9}}$, 
I.~Felis$\mathrm{^{2}}$,
M.~Frullini$\mathrm{^{5}}$,
W.~Fulgione$\mathrm{^{3,10,a}}$, 
G.~Lucchini$\mathrm{^{1,7}}$,
L.~Manara$\mathrm{^{1,7}}$, 
M.~Maspero$\mathrm{^{1,11}}$,
A.~Papagni$\mathrm{^{1,12}}$,
M.~Perego$\mathrm{^{1,3}}$,
R.~Podviyanuk$\mathrm{^{3}}$,
A.~Pullia$\mathrm{^{1,7,*}}$,
A.~Quintino$\mathrm{^{5,8}}$,
N.~Redaelli$\mathrm{^{1}}$,
E.~Ricci$\mathrm{^{5}}$,
A.~Santagata$\mathrm{^{4}}$,
D.~Sorrenti$\mathrm{^{1,6}}$,
L.~Zanotti$\mathrm{^{1,7}}$.\\
\\

\small{
1) INFN, Sezione di Milano Bicocca, P.za della Scienza 3, I-20135 Milano, Italy

2) Universitat Polit\`ecnica de Val\`encia Camino de Vera, s/n 46022 Valencia, Spain

3) INFN, LNGS, Via G. Acitelli 22, I-67100 Assergi (L'Aquila), Italy

4) ENEA, Centro Ricerche Casaccia, Via Anguillarese 301, I-00123 S. Maria di Galeria (Roma), Italy

5) DIAEE, Sapienza Universit\`a di Roma, Via Eudossiana 18, I-00184 Roma, Italy

6) Dip.Inform., Sist. e Comunic., Universit\`a di Milano Bicocca, viale Sarca 336, I-20126 Milano, Italy

7) Dip. di Fisica, Universit\`a di Milano Bicocca, P.za della Scienza 3, I-20126 Milano, Italy

8) INFN, Sezione di Roma La Sapienza, P.le Aldo Moro 2, I-00185 Roma, Italy

9) CERN, CH-1211 Geneva 23, Switzerland

10) INAF, Osservatorio Astrofisico di Torino, I-10025 Pino Torinese (Torino), Italy

11) DISAT, Universit\`a di Milano Bicocca, P.za della Scienza 3, I-20126 Milano, Italy

12) Dip. di Scienza dei Materiali, Universit\`a di Milano Bicocca, I-20126 Milano, Italy
}\\



\small{
$\mathrm{*)}$  Spokesperson

$\mathrm{a)}$  walter.fulgione@lngs.infn.it

$\mathrm{b)}$  massimo.corcione@uniroma1.it
}
\\

\begin{abstract}
The MOSCAB experiment (Materia OSCura A Bolle) uses the "geyser technique", a variant of the superheated liquid technique of extreme simplicity. 
Operating principles of the new dark matter detector and technical solutions of the device are reported in detail.
First results obtained in a series of test runs taken in laboratory demonstrate that we have successfully built and tested a geyser-concept bubble chamber that can be used in particle physics, especially in dark matter searches, and that we are ready to move underground for extensive data taking.
\end{abstract}
\section{Introduction}
\label{intro}
The existence of non-baryonic, non-relativistic dark matter (DM) is by now well established \cite{a}, yet the nature of particle dark matter remains one of the most important quests in the field of particle physics \cite{b}. 
Among candidates for non-baryonic DM, weakly interacting massive particles (WIMPs) provide solutions for outstanding issues in both cosmology and particle physics. In this framework, a number of experiments have been designed and are operating with the aim to discover the WIMP particle by observing its collisions with ordinary matter \cite{c}.\\
The sensitivity of a dark matter direct detection experiment depends on the product between the local WIMP flux, that for a fixed local density depends on the WIMP mass, and the interaction cross section.
This is why the experimental observables are expressed in the WIMP mass-cross section plane.
The cross section depends on the nature of the WIMP-nucleon coupling, including spin-dependent WIMP-nucleon and spin-independent interactions. 
For spin-dependent coupling, the cross section depends on the nuclear spin factor. 
The $\mathrm{^{19}F}$ nucleus, because of its single unpaired proton and 100\% isotopic abundance, provides a unique target to search for the spin-dependent WIMP-proton interactions \cite{g}.
Experiments using fluorine-based targets are setting the strongest constraints on such interactions \cite{d,e,ea} and all of them make use of bubble chambers \cite{ee,iiii}, i.e., detectors containing superheated liquids, whose main advantage is that of being insensitive to beta and gamma background.
The aim of such detectors, for the best possible sensitivity, is to have the largest possible target mass and the lowest possible detection threshold.\\
%
Our detector, whose main features will be discussed in detail in the next section, is a geyser-concept bubble chamber \cite{h,i} that we plan to employ to explore the spin-dependent WIMP-proton coupling. 
\section{The MOSCAB bubble chamber}
\label{sec:1}
\subsection{Fundamentals of the geyser technique}

In the geyser-concept bubble chamber a closed vessel is filled with a target liquid (e.g., $\mathrm{C_3F_8}$) and its saturated vapour, separated by a layer of buffer liquid (e.g., propylene glycol $\mathrm{ C_3 H_8 O_2}$). 
The bottom part of the vessel, which contains the target liquid, is kept in a thermal bath at constant temperature $\mathrm{T_{liq}}$. The saturated vapour above it is kept at a constant temperature $\mathrm{T_{vap}}$, such that $\mathrm{T_{vap} < T_{liq}}$.
As the vapour pressure $\mathrm{p_{vap}(T_{vap})}$ inside the vessel is lower than the vapour pressure $\mathrm{p_{vap}(T_{liq})}$, the target liquid is in underpressure of $\mathrm{\Delta p = p_{vap}(T_{liq}) - p_{vap}(T_{vap})}$, and therefore in a superheated metastable state. 
The degree of metastability of the target liquid at temperature $\mathrm{T_{liq}}$ can also be expressed as:
$\mathrm{\Delta T=T_{liq}}$
$\mathrm{-T_{vap}}$.
It should be emphasised that the buffer liquid, which avoids surface evaporation and has important insulation purposes, is indispensable to reach high superheats. \\
%
%
%
When an incoming particle, e.g., a WIMP, interacts with a target nucleus, should the energy released by the recoiling ion be higher than a given threshold energy, $\mathrm{E_{thr}}$, and deposited within a certain critical volume, a vapour bubble is produced. This vapour bubble rises through the superheated liquid pushing up some of this liquid to cross the interface with the overlying buffer liquid, which is the reason for the name geyser.
After the condensation of the excess vapour occurs in the cooled top portion of the vessel, the drops of condensate fall back into the target liquid, which consequently recovers its original superheated state.\\  
Accordingly, a geyser-concept bubble chamber has a number of advantages compared with a standard bubble chamber.
In fact, its operation is based on a simple and continuous process of evaporation and condensation, such that, after each event, the detector-reset to the initial state is automatic.
Moreover, the absence of moving parts eliminates any possible particulate detachment (possibly occurring during the compression and re-expansion operations required in a standard bubble chamber), thus avoiding that such impurities may act as nucleation sites.
\subsection{Detector's main features}
\begin{figure}
\centering 
\vspace{-2.cm}       
\resizebox{1.0\textwidth}{!}{%
  \includegraphics{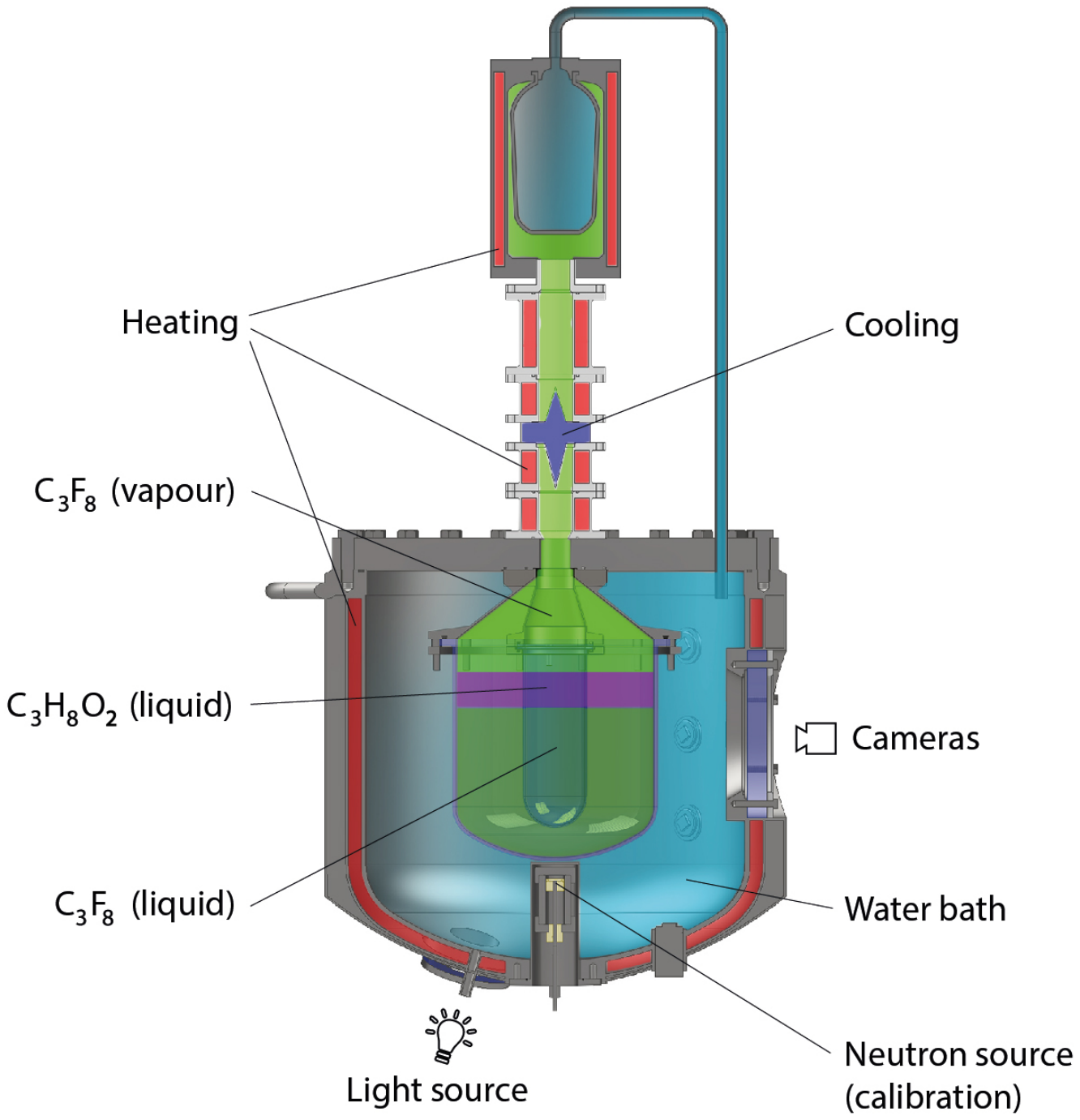}
}
\vspace{-2.0cm}       
\caption{The MOSCAB bubble chamber.}
\label{fig.1}       
\end{figure}

The MOSCAB (Materia OSCura A Bolle) bubble chamber consists of three main parts symbolically named as the "body", the "neck" and the "head", mounted one above the other, as shown in the schematic displayed in Figure \ref{fig.1}.
\\
The core of the body is a rounded-bottom, cylindrical quartz vessel 
which contains liquid $\mathrm{C_3F_8}$, overlaid by a propylene glycol buffer which separates the target liquid from its saturated vapour, maintained at a lower temperature. 
The detector can be equipped with either a 2L vessel or a 25L vessel.
The 2L vessel has a wall-thickness of 3 mm, an internal diameter of 96 mm and an overall height of 284 mm.
The 25L vessel has a wall-thickness of 5 mm, an internal diameter of 314 mm and an overall height of 345 mm.
The quartz vessel is centred within a stainless-steel, jacketed pressure-tank, filled with water providing both radiation shielding and temperature control of the target liquid. The pressure-tank is a rounded-bottom cylinder with an internal diameter of 600 mm and a total height of 615 mm. The water bath temperature is regulated by means of a water-ethylene glycol mixture continuously flowing through the tank jacket, pumped from a recirculating LAUDA Integral XT 150 air-cooled thermostat.
The tank jacket is equipped with the hydraulic connections necessary for filling, unloading and bleeding operations, and with a safety valve for protection against overpressure. 
Additionally, a pair of viewports located on the side of the tank, and positioned at a stereo angle of 90-deg, are used to monitor the chamber, whereas a pair of smaller viewports located at the bottom of the tank are used for the interior illumination by LED lamps.
A hole with a bolted cap, drilled in the middle of the bottom of the tank, allows the positioning of the holder of the neutron calibration source just below the quartz vessel. 
The top-cover of the tank includes six openings with removable plugs for the bleeding of air after water-filling; moreover, its internal side bears a stainless-steel, conical flange protruding downwards in the axial direction to support the quartz vessel, attached to it by means of bolts, using an elastomer O-ring to seal the quartz to the steel flange.
In particular, the support-flange for the 2L vessel has an overall height of 120 mm with an opening angle of 17-deg, whereas that for the 25L vessel has an overall height of 112 mm with an opening angle of 52-deg.
A hole perforated in the centre of the top-cover allows communication between the inside of the quartz vessel and the overhead neck, which is filled by the saturated vapour.\\
The neck basically consists of a 54-mm-diameter stainless-steel tube with jackets around it, in which the water-ethylene glycol mixture pumped from the LAUDA Integral XT 150 air-cooled thermostat cited earlier continuously flows for temperature control. 
The neck is subdivided into three parts, superimposed one above the other. 
The lower part contains the so-called "condensation bulb", consisting of a shell formed of two vertically-aligned, stainless-steel cones soldered at their base, suspended in the centre of the tube to leave a free passage for the vapour up through the neck.
A pair of opposite hydraulic connections at the sides of the double-cone allow for the continuous flow of a cold water-ethylene glycol mixture pumped from a dedicated recirculating LAUDA RP 845 air-cooled thermostat, to keep constant the condensation temperature. 
The middle part of the neck contains a filter to avoid particle contamination from the $\mathrm{C_3F_8}$ filling equipment located above.
Finally, the upper part holds a pair of lateral connection ports to the $\mathrm{C_3F_8}$ charge line, and a top flange for the support of the head.\\
The head is a pressure equaliser consisting of a stainless-steel, jacketed cylinder, divided by a rubber membrane into a bottom part, which contains the saturated vapour, and an upper part, connected to the tank via a pipe, which contains water. The jacket temperature is controlled by the same water-ethylene glycol mixture pumped from the LAUDA Integral XT 150 thermostat. 
A safety valve for protection against overpressure inside the vessel is mounted by the side of the saturated vapour.

\subsection{Equipment and instrumentation}
The apparatus is equipped with two cameras to photograph the chamber, two hydrophones to record the acoustic emissions from bubble nucleation, as well as seven temperature detectors and two pressure transducers to monitor the thermodynamic parameters.\\
The cameras, firmly attached to the side viewports, are black-and-white Basler avA1600-50gm GigE digital cameras, delivering 55 frames per second at a resolution of 1600x1200 pixels, each equipped with a 8.25 mm high resolution Goyo GMTHR38014MCN optic.\\
The hydrophones, tightly attached to the conical flange and suspended in the water bath in the close proximity of the outer wall of the quartz vessel, are Co.l.mar. AR0190XS omni-directional hydrophones, with a receiving sensitivity of -208 dB re $\mathrm{1V/\mu Pa}$ @ 50 kHz without preamplifier, and a working band up to 170 kHz.\\
The temperature detectors are Tersid PT100-4 thermoresistances with a measurement accuracy of $\mathrm{0.05~^oC}$: five of them are placed in the water tank to measure the bath temperature at different places, and two are located outside the detector to measure the room temperature.
The pressure transducers are Micro Sensor MPM4760 piezoelectric pressure transmitters with a full scale accuracy of $\mathrm{\pm 0.1\%}$ (f.s.=16 bar): one is placed in the water tank, the other inside the neck, just above the condensation bulb.\\
The data acquisition system is a National Instruments cRIO-9075 embedded FPGA controller equipped with three NI 9217 expansion boards for the temperature data processing and a NI 9207 expansion board for the pressure data processing.

\section{Data taking}
\label{sec:2}
\begin{figure}
\centering 
\hspace*{-2.5cm}
\vspace{-0.5cm}
\resizebox{1.4\textwidth}{!}{%
  \includegraphics{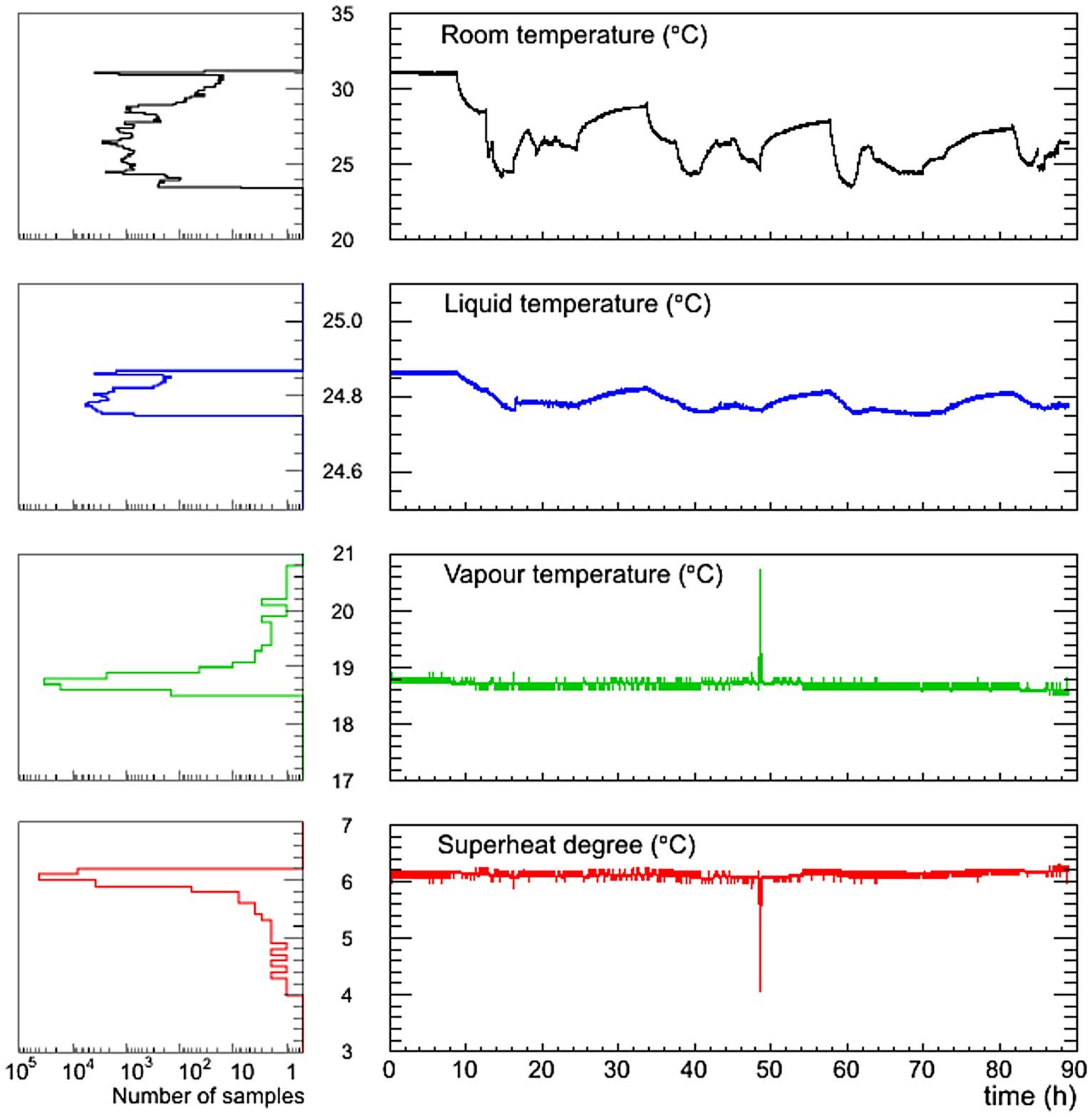}
}
\vspace{0.5cm}       
\caption{Time-variations of the thermodynamic parameters during the 90 hour-long run at $\mathrm{\Delta T\cong 6^oC}$, with their distributions (from the top: room temperature; water bath temperature; saturated vapour temperature; superheat degree).}
\label{fig.2}       
\end{figure}
Many tests have been performed above ground in the Milano Bicocca INFN Laboratory, in view of moving the detector underground to the INFN Gran Sasso Laboratory.
We examined various temperatures of the liquid $\mathrm{C_3F_8}$ before deciding to operate mainly at 25$\mathrm{^o C}$, at which the detector shows an excellent stability even if stressed by high nucleation rates due to the cosmic ray neutron background, and to an AmBe 40 kBq neutron source, when applied.
At this temperature we carried out several runs using 1 kg of liquid $\mathrm{C_3F_8}$, which corresponds to a sensitive volume of 0.75 litres, kept at different superheat degrees ranging from 4$\mathrm{^o C}$ to 9$\mathrm{^o C}$.
During each run, having a typical duration of several hours, the thermodynamic parameters are recorded every 6 s, whereas the target liquid inside the vessel is continuously camera-monitored by taking one picture every 18 ms.
Any change appearing between two consecutive pictures triggers the storage of a set of images that will be processed later.\\
As an example, the data taken during a 90 hour-long run at $\mathrm{\Delta T \cong 6^oC}$ are displayed in Figure \ref{fig.2}, in which the room temperature, the water bath temperature, the saturated vapour temperature derived from the vapour pressure \cite{f}, and the superheat degree, are plotted versus time, together with their distributions.
The corresponding working pressure, $\mathrm{p_{vap}(T_{vap})}$, is 730.1 kPaA, whereas the vapour pressure at the liquid temperature, $\mathrm{p_{vap}(T_{liq})}$, is 862.8 kPaA, which means that the target liquid is in underpressure of $\mathrm{\Delta p = 132.7}$ kPaA.\\
\begin{figure}
\vspace{-1.5cm}
\centering 
\resizebox{1.0\textwidth}{!}{%
  \includegraphics{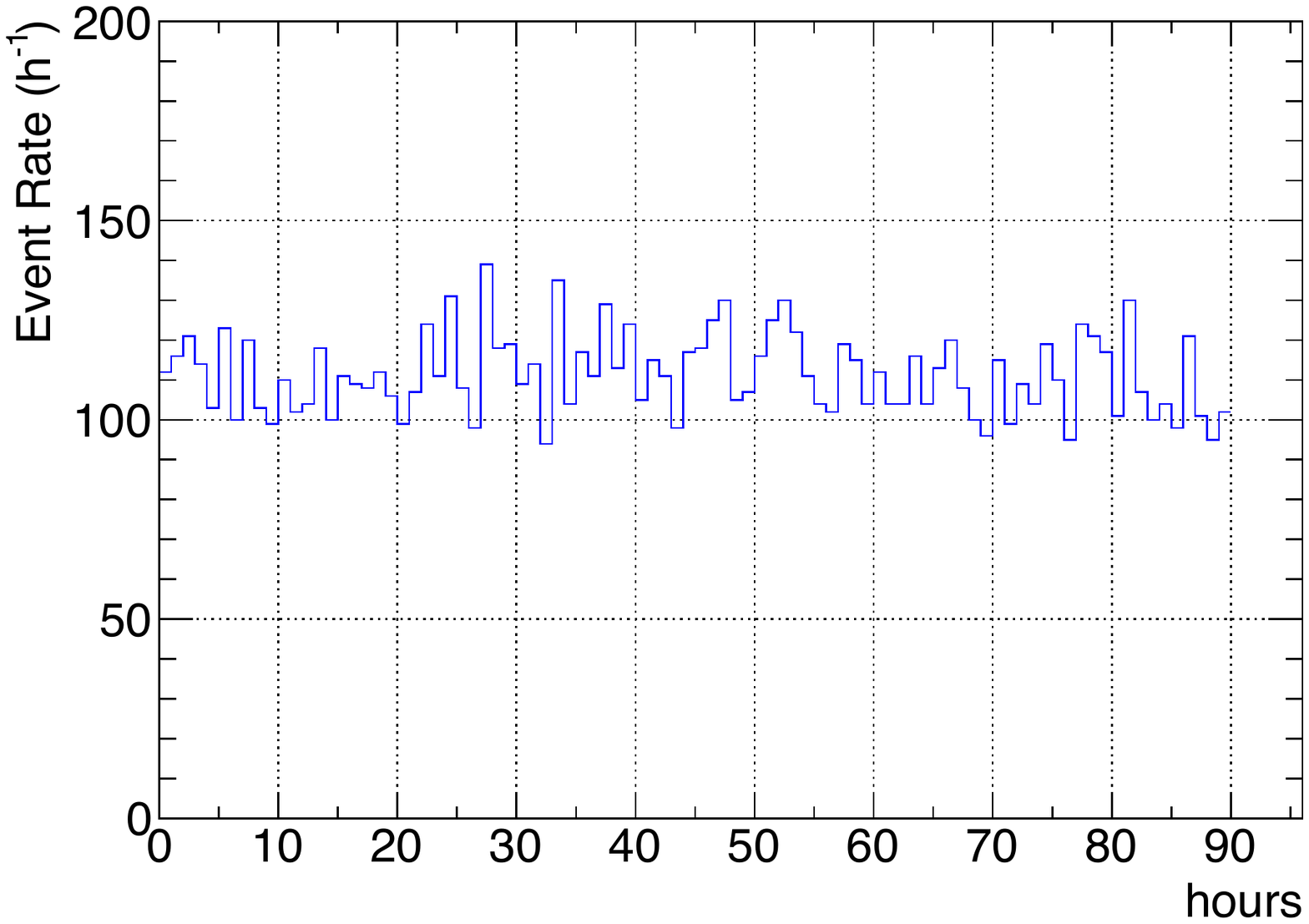}
}
\vspace{-1.5cm}
\caption{Time-variations of the event rate during the 90 hours-long run at $\mathrm{\Delta T \cong 6^oC}$.}
\label{fig.3}       
\end{figure}
The related bubble counting rate is reported in Figure \ref{fig.3}, whereas the distribution of the time intervals between consecutive events is shown in Figure \ref{fig.4}, proving the randomness of the arrival time of the events. A dead time smaller than 3 s can be inferred, as a consequence of the time required to store the set of frames. Same durations are observed for many other examined test runs, demonstrating that the data acquisition dead time is independent of the event rate.
\begin{figure}
\vspace{-1.5cm}
\centering 
\resizebox{1.0\textwidth}{!}{%
  \includegraphics{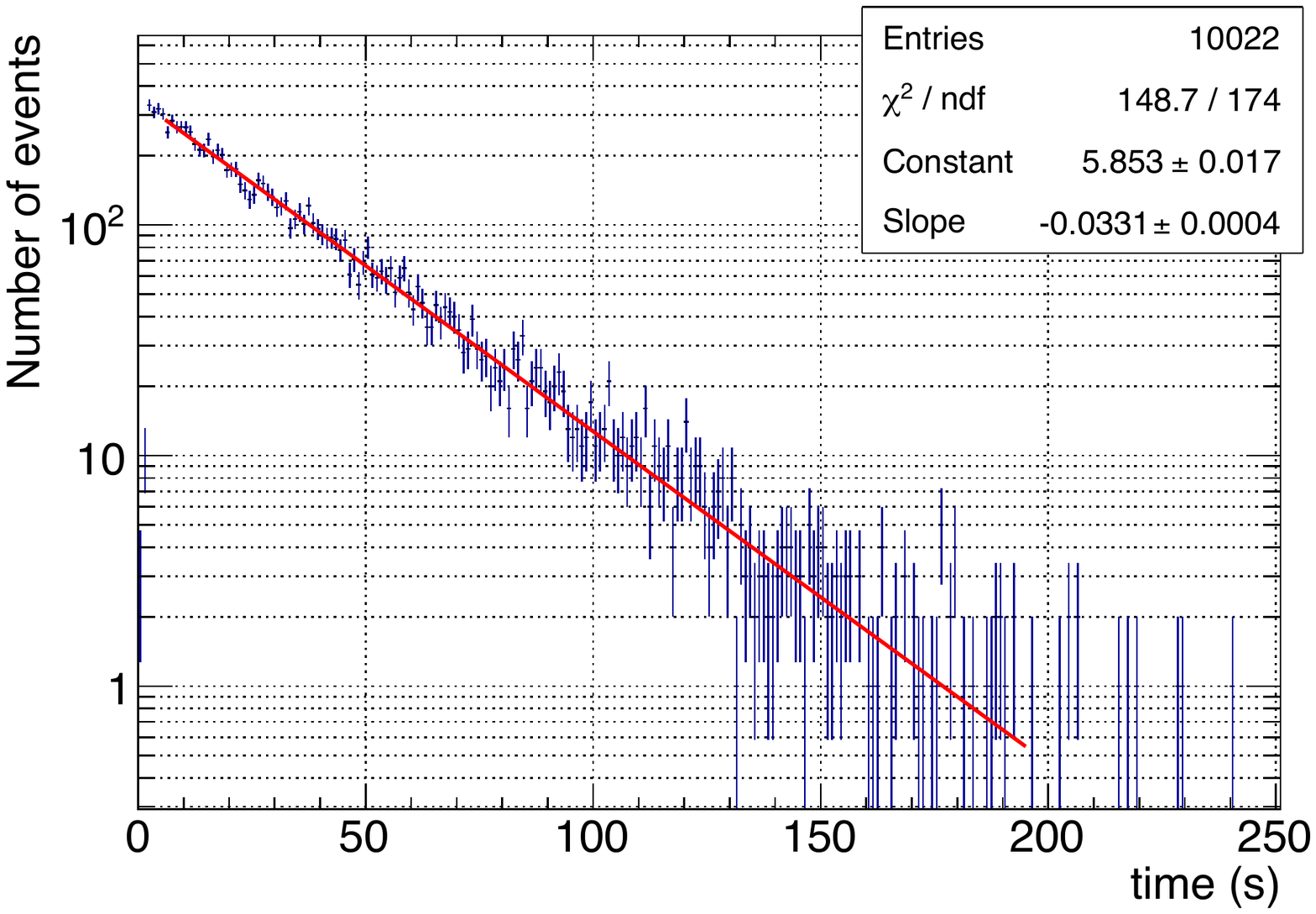}
}
\vspace{-1.9cm}
\caption{Distribution of the time intervals between consecutive events occurred during the 90 hour-long run at $\mathrm{\Delta T\cong 6^o C}$.}
\label{fig.4}       
\end{figure}
It is apparent that, despite the pronounced shuffling of the room temperature, the temperature of the water bath remains stable around the value of $\mathrm{24.8^oC}$, with changes below $\mathrm{0.1^oC}$ that follow quite closely the time variations of the  room temperature.
Also the overall time evolution of the superheat degree remains stable around the value of $\mathrm{6.06^oC}$, upon which the fluctuations of the temperature difference due to the formation of bubbles and their subsequent condensation are clearly superimposed.
Moreover, the time distribution of the superheat degree is marked by a negative pulse of $\mathrm{\sim 2^oC}$ of amplitude, due to an abrupt bubble boiling initiated at the vessel wall by a gas pocket acting as a nucleation site. 
A close-up of this pulse is shown in Figure \ref{fig.5}, in which the recovery of the original superheat degree is clearly seen, the complete recovery time being about 6 minutes. 
It is important to stress that during the whole process the device is still sensitive, but at a lower $\mathrm{\Delta T}$.
\begin{figure}
\vspace{-1.cm}
\centering 
\resizebox{0.67\textwidth}{!}{%
  \includegraphics{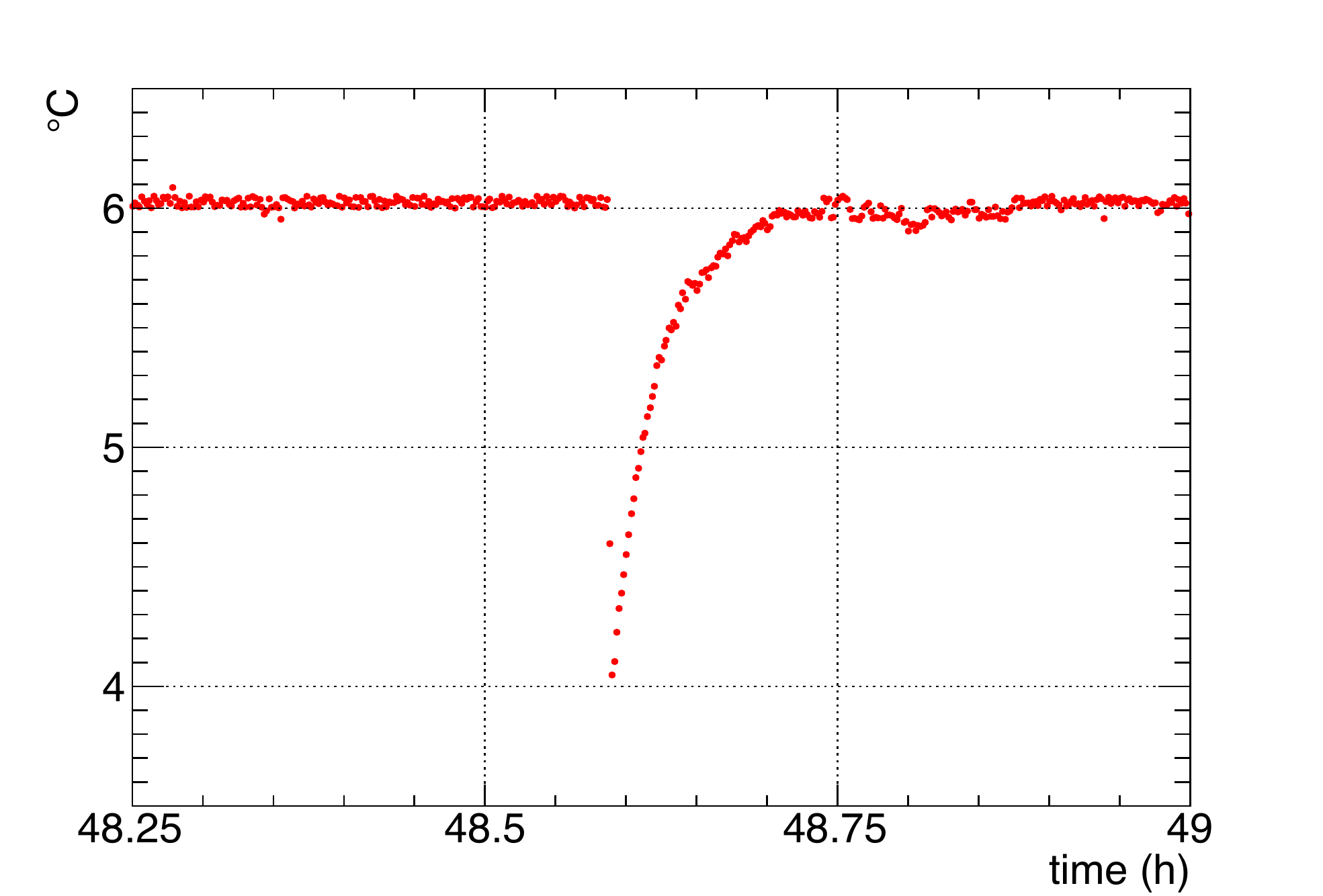}
}
\caption{Time-evolution of the superheat degree along the sudden bubble nucleation occurred at the vessel wall during the 90 hour-long run at $\mathrm{\Delta T \cong 6^o C}$.}
\label{fig.5}       
\end{figure}\\
The above behaviour can be described by the large vapour flow rate reaching the top and producing a pressure spike which drastically reduces the applied superheat. 
On a longer time scale the returning recondensed liquid slightly cools the main liquid volume, also reducing the superheat. 
However, on one side, the decrease of the superheat degree inhibits the bubble boiling at the wall and, on the other side, the continuous condensation of the excess vapour occurring at the top reduces the pressure, which brings the detector back to the original settings.
As stated above, the propylene glycol layer seems to play a critical role in the working of the geyser-concept bubble chamber, 
and further tests on this point must be carried out in view of increasing the sensitive volume.
Notice that in some situations the vapour flow rate produced by the continuous boiling at the vessel wall can be so large as to destroy the superheat, leading the detector to stop working instead of recovering its initial metastable state.  
This phenomenon is greater at higher superheats and event rates. 
The problem is still under investigation, and a great improvement has been obtained by using flame polished quartz, whose internal surface finish plays a crucial role. 
Any time the superheated state of the target liquid is destroyed due to wall events, a procedure which takes about 5 hours has been found which resets the detector. 
First, the circulation of the cold water-ethylene glycol mixture through the condensation bulb is stopped. The top part of the neck is then cooled to produce condensation and accumulate the condensed liquid above the filter. Subsequently, the top is heated to increase the vapour pressure and force the liquid back into the main volume. Next, the temperature at the top is equalised to the temperature of the thermal bath. Once the level of liquid $\mathrm{C_3F_8}$ inside the quartz vessel is restored, the detector is restarted by reactivating the circulation through the condensation bulb to put the target liquid in underpressure.
\\
\section{First results and discussion} 
\label{sec:2}
\begin{figure*}
\centering 
\vspace{-3.5cm}       
\hspace*{-1.5cm}       
\resizebox{1.2\textwidth}{!}{%
  \includegraphics{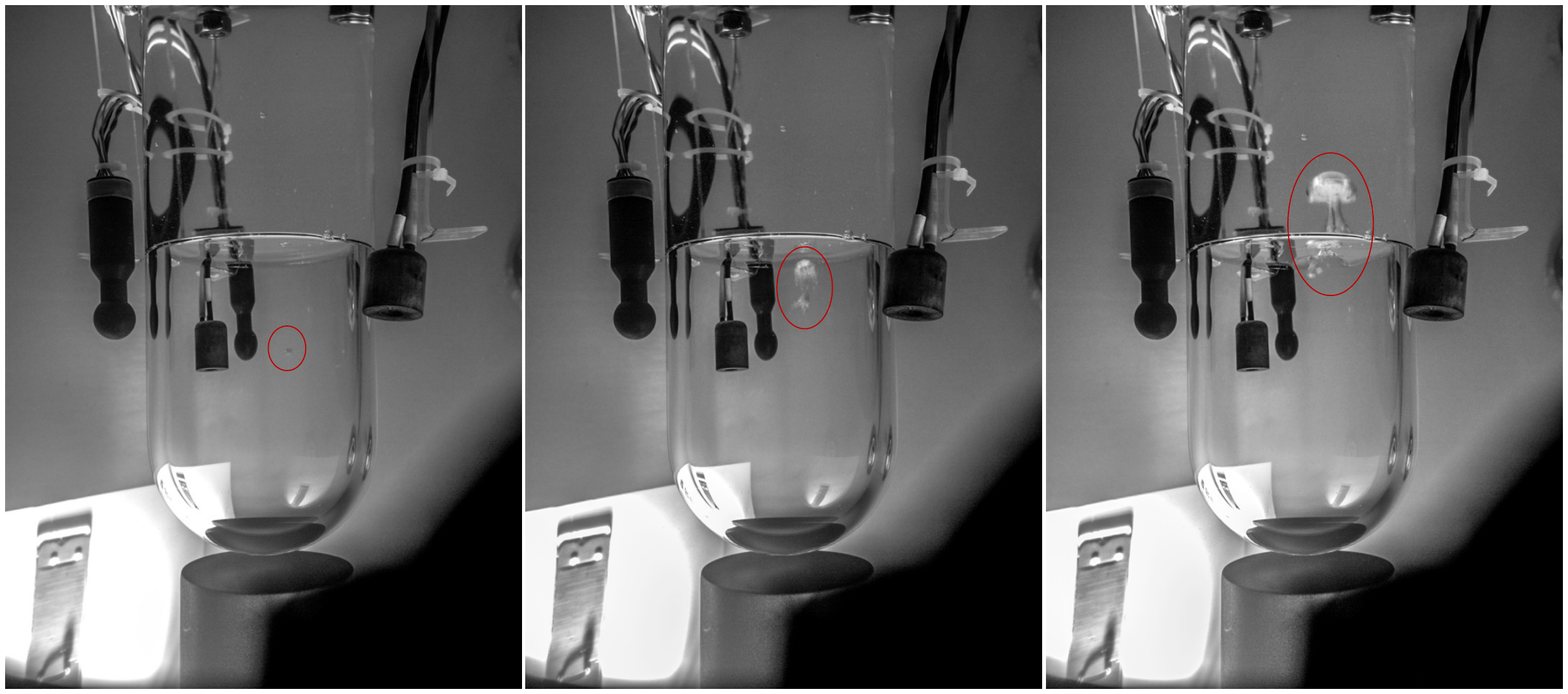}
}
\vspace{-4.5cm}       
\caption{A typical bubble nucleation and geyser formation occurred during a run at $\mathrm{\Delta T = 8.47^oC}$.}
\label{fig.6}       
\end{figure*}

\begin{figure*}
\centering 
\vspace{-2.0cm}       
\hspace*{-2.5cm}       
\resizebox{1.3\textwidth}{!}{%
\includegraphics{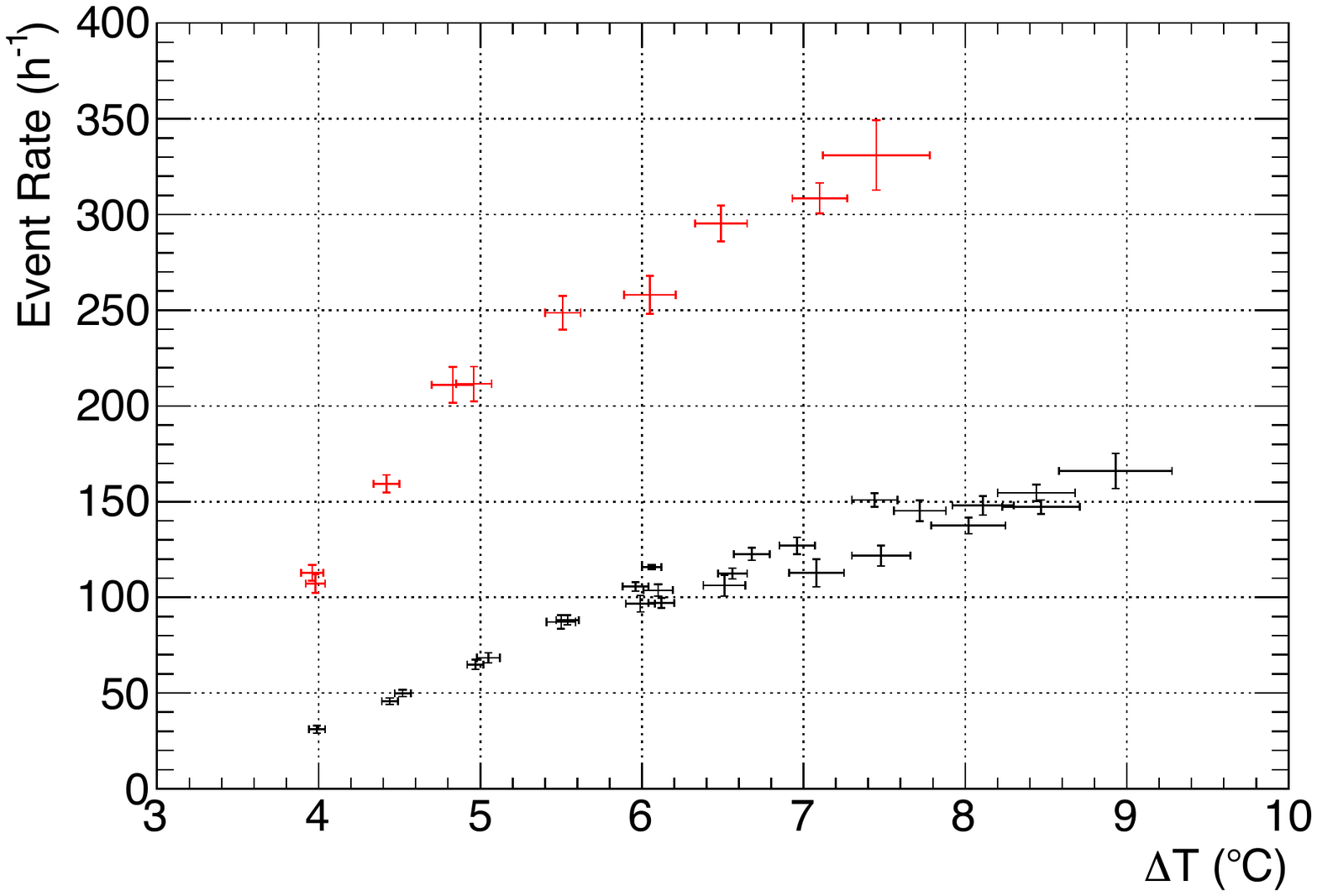}
}
\vspace{-2.5cm}       
\caption{Event rate versus superheat degree, with and without the 40 kBq AmBe neutron source.
Horizontal error bars represent the standard deviation of the $\mathrm{\Delta T}$ distribution during the run.}
\label{fig.7}       
\end{figure*}

In May and June 2017, we have carried out 56 test runs, to check the overall functioning of the 2L scale-up of the MOSCAB detector for a total live time of about 400 hours. 
As said earlier, the detector has been tested at various superheat degrees between $\mathrm{4^oC}$ and $\mathrm{9^oC}$, which corresponds to a range of recoil threshold energies of the order of 100 keV - 10 keV, with and without a 40 kBq AmBe neutron source placed just below the quartz vessel.
A typical bubble nucleation and the related geyser formation at the interface between the target liquid and the buffer liquid, occurred during a run executed at $\mathrm{\Delta T= 8.47^o C}$ with only background radiation, is shown in the sequence reported in Figure \ref{fig.6}. 
Notice that the main source of background, for a detector above ground, insensitive to gammas and electrons and surrounded by 25 cm of water that thermalises with good efficiency the neutrons from external radioactivity, is the secondary cosmic ray radiation.
\\
The results obtained are reported in Figure \ref{fig.7}, where the counting rate is plotted versus the imposed superheat degree. 
As expected, the detector is sensitive to the neutrons emitted by the AmBe source. 
Moreover, the event rate with and without the neutron source increases as superheat is increased. 
The same results are also listed in Table \ref{tab:1}, in which the average of the superheat distribution obtained by measuring the temperature difference every 6 seconds and the corresponding standard deviation (std dev), the time duration, the number of recorded events, and the average of the event rate distribution obtained from the recorded hourly number of triggers, are listed for each performed test run. 
It is apparent that, even in the most severe conditions, i.e., at high $\mathrm{\Delta T}$ and/or high nucleation rates due to the presence of the neutron source, the thermodynamic conditions remain reasonably stable. 
Notice that, with the scope to verify if each trigger actually corresponds to a real physical event, the captured images relevant to more than 2000 events have been individually scanned by eye, which led us to conclude that the counting rate based on the output of the automatic triggering system is reliable.
\\
Interestingly, according to the thermal spike theoretical approach \cite{ee,iiii}, at the largest superheat degree reached in our tests, i.e., $\mathrm{\Delta T \cong 9^oC}$, the minimum energy release required for a bubble formation should be of the order of 10 keV, which means that now we are ready to carry out underground measurements of the actual energy threshold of our detector with the help of a $\mathrm{^{88}YBe}$ neutron source \cite{ccc}.
In this connection, it seems worth pointing out that in the present experimental conditions above ground, in which the cosmic ray neutron background plays an important role, neither a fiducial volume inside the target liquid has been defined by a precise determination of the position of the bubble nucleation, nor the discrimination between alpha and neutron bubble nucleation has been realised.
Precise measurements of the detector's internal and external background, as well as the determination of its sensitivity and capability to reject alpha background, will be carried out in the forthcoming underground tests, where the neutron background from cosmic rays is almost absent. 
\begin{table}
\caption{ {\bf{ MOSCAB data taking above ground in the Milano Bicocca INFN Laboratory at $\mathrm{T_{L}=25^o}$C  }}}
\label{tab:1}
\begin{tabular*}{\columnwidth}{@{\extracolsep{\fill}}lrrrrl@{}}
\\
\hline
\\
RUN n. & $\mathrm{<\Delta T>}$ $\mathrm{[^oC]}$& std dev $\mathrm{[^oC]}$ & duration [h]& $\mathrm{N_{ev}}$ &  $\mathrm{<R_{ev}>}$ $\mathrm{[h^{-1}]}$\\
\\
\hline
\\
1180-bk & $\mathrm{3.99}$ & $\mathrm{0.05}$ & 15.1 & 468 & $\mathrm{31.1}\pm2.0$\\
1181-bk & $\mathrm{4.52}$ & $\mathrm{0.05}$ & 14.7 & 721 & $\mathrm{49.9}\pm1.8$\\
1198-bk & $\mathrm{4.44}$ & $\mathrm{0.05}$ & 14.5 & 651 & $\mathrm{45.7}\pm1.7$ \\
1199-bk & $\mathrm{4.97}$ & $\mathrm{0.05}$ & 18.4 & 1197 & $\mathrm{64.9}\pm2.5$ \\
1182-bk & $\mathrm{5.05}$ & $\mathrm{0.07}$ & 9.8 & 675 & $\mathrm{68.4}\pm2.6$ \\
1183-bk & $\mathrm{5.50}$ & $\mathrm{0.09}$ & 7.0 & 600 & $\mathrm{87.2}\pm3.6$ \\
1200-bk & $\mathrm{5.54}$ & $\mathrm{0.07}$ & 13.7 & 1156 & $\mathrm{88.1}\pm2.5$ \\
1201-bk & $\mathrm{5.96}$ & $\mathrm{0.08}$ & 13.0 & 1383 & $\mathrm{105.6}\pm2.3$ \\
1178-bk & $\mathrm{5.99}$ & $\mathrm{0.09}$ & 9.2 & 891 & $\mathrm{96.7}\pm4.3$ \\
1232-bk & $\mathrm{6.06}$ & $\mathrm{0.06}$ & 90.0 & 10023 & $\mathrm{115.8\pm 1.1}$ \\
1228-bk & $\mathrm{6.10}$ & $\mathrm{0.09}$ & 10.0 & 1038 & $\mathrm{103.7}\pm3.1$\\
1204-bk & $\mathrm{6.12}$ & $\mathrm{0.08}$ & 21.8 & 2128 & $\mathrm{97.1}\pm2.7$ \\
1184-bk & $\mathrm{6.51}$ & $\mathrm{0.13}$ & 5.0 & 527 & $\mathrm{106.2}\pm5.6$ \\
1202-bk & $\mathrm{6.56}$ & $\mathrm{0.09}$ & 14.9 & 1672 & $\mathrm{112.5}\pm2.8$ \\
1205-bk & $\mathrm{6.68}$ & $\mathrm{0.11}$ & 10.3 & 1273 & $\mathrm{122.6}\pm3.3$ \\
1203-bk & $\mathrm{6.96}$ & $\mathrm{0.11}$ & 10.2 & 1289 & $\mathrm{127.0}\pm4.4$ \\
1189-bk & $\mathrm{7.08}$ & $\mathrm{0.17}$ & 4.4 & 501 & $\mathrm{112.7}\pm7.2$ \\
1208-bk & $\mathrm{7.44}$ & $\mathrm{0.14}$ & 8.9 & 1330 & $\mathrm{150.9}\pm3.6$ \\
1190-bk & $\mathrm{7.48}$ & $\mathrm{0.18}$ & 4.2 & 520 & $\mathrm{121.7}\pm5.4$ \\
1207-bk & $\mathrm{7.72}$ & $\mathrm{0.16}$ & 8.4 & 1204 & $\mathrm{145.2}\pm5.5$ \\
1179-bk & $\mathrm{8.02}$ & $\mathrm{0.23}$ & 4.3 & 597 & $\mathrm{137.5}\pm4.2$ \\
1206-bk & $\mathrm{8.11}$ & $\mathrm{0.19}$ & 9.2 & 1359 & $\mathrm{148.0}\pm5.0$ \\
1209-bk & $\mathrm{8.44}$ & $\mathrm{0.24}$ & 6.0 & 928 & $\mathrm{154.6}\pm4.4$ \\
1191-bk & $\mathrm{8.47}$ & $\mathrm{0.24}$ & 6.2 & 916 & $\mathrm{147.2}\pm3.7$ \\
1233-bk & $\mathrm{8.93}$ & $\mathrm{0.35}$ & 2.8 & 431 & $\mathrm{166.0}\pm9.2$ \\
\\
\hline
\\
1212-AmBe & $\mathrm{3.96}$ & $\mathrm{0.07}$ & 6.5 & 708 & $\mathrm{112.8}\pm4.1$ \\
1224-AmBe & $\mathrm{3.98}$ & $\mathrm{0.06}$ & 8.6 & 919 & $\mathrm{107.2}\pm4.9$ \\
1225-AmBe & $\mathrm{4.42}$ & $\mathrm{0.08}$ & 7.4 & 1168 & $\mathrm{159.3}\pm4.6$ \\
1211-AmBe & $\mathrm{4.83}$ & $\mathrm{0.13}$& 2.4 & 495 & $\mathrm{211.0}\pm9.3$ \\
1215-AmBe & $\mathrm{4.96}$ & $\mathrm{0.11}$ & 4.1 & 857 & $\mathrm{211.5}\pm9.1$ \\
1216-AmBe & $\mathrm{5.51}$ & $\mathrm{0.11}$ & 3.2 & 803 & $\mathrm{248.7}\pm8.8$ \\
1219-AmBe & $\mathrm{6.05}$ & $\mathrm{0.16}$ & 2.7 & 721 & $\mathrm{258.0}\pm9.9$ \\
1220-AmBe & $\mathrm{6.49}$ & $\mathrm{0.16}$ & 3.5 & 1054 & $\mathrm{295.3}\pm9.3$ \\
1226-AmBe & $\mathrm{7.10}$ & $\mathrm{0.17}$ & 4.8 & 1485 & $\mathrm{308.5}\pm8.0$ \\
1227-AmBe & $\mathrm{7.45}$ & $\mathrm{0.33}$ & 1.0 & 331 & $\mathrm{331.0}\pm18.2$ \\
\\
\hline
\end{tabular*}
\end{table}
%
%
%
\section{Conclusions}
The results obtained show that we have successfully built and tested a geyser-concept bubble chamber that can be used in particle physics, in particular for spin-dependent WIMP-proton interacting dark matter direct search. We are now ready to move the detector underground to carry out further tests concerning improved optics, the volume of target liquid, the height of the buffer layer, and the cooling configuration of the neck, all important for the necessary sensitive volume increases and understanding at depth the device.
Clearly, acoustic detection for background reduction and a detailed study of energy threshold behaviour using radioactive sources would be high on the program.

\section{acknowledgements}
We  gratefully acknowledge support from  Istituto Nazionale di Fisica Nucleare (INFN) and Spanish Ministerio de Econom\`ia y Competitividad (MINECO), Grants FPA-2015-65150-C3-2-P
(MINE-CO/FEDER), and Consolider MultiDark CSD2009-00064. 
We are grateful to Laboratori Nazionali del Gran Sasso (LNGS) for hosting and supporting the MOSCAB project.
We thank Massimo Locatelli, Tecnologia Meccanica, Treviolo (BG), Italy, for supporting our project with hard work, competence and passion.



\end{document}